# How Much Should I Pay? An Empirical Analysis on Monetary Prize in TopCoder

Mostaan Lotfalian Saremi, Razieh Saremi, Denisse Martinez-Mejorado

Stevens Institute of Technology, Hoboken NJ 070390, USA
{mlotfali, rsaremi, dmartin2}@stevens.edu

**Abstract.** It is reported that task monetary prize is one of the most important motivating factors to attract crowd workers. While using expert-based methods to price Crowdsourcing tasks is a common practice, the challenge of validating the associated prices across different tasks is a constant issue. To address this issue, three different classifications of multiple linear regression, logistic regression, and K-nearest neighbor were compared to find the most accurate predicted price, using a dataset from TopCoder website. The result of comparing chosen algorithms showed that the logistics regression model will provide the highest accuracy of 90% to predict the associated price to tasks and KNN ranked the second with an accuracy of 64% for K = 7. Also, applying PCA wouldn't lead to any better prediction accuracy as data components are not correlated.

## 1 Introduction

Available literatures on motivation patterns of crowdsourcing workers have reported that the monetary prize associated with tasks is one of the top motivating factors to attract and involve potential workers in task competition [1]. The monetary prize usually represents the degree of task complexity as well as required competition levels [2][3]. In practice, task requesters frequently employ expert-based methods to price tasks, which may involve a high degree of subjectivity, while the challenge of validating the associated prices across different tasks is a constant issue.

As of today, multiple pricing models are introduced with a focus on pricing strategy [4], the Context-Centric Pricing approach [ 5], the impact of price on workers' behavior [6] and using machine learning methods [ 7] to help task requesters with predicting reasonable price range. However, none of these papers applied PCA to evaluate the accuracy of the presented. In this work, we aim to investigate that gap.

To address this issue, three different classifications of multiple linear regression, logistic regression, and K-nearest neighbor were compared to find the most accurate predicted price, using data extracted from TopCoder website [8]. The result of comparing chosen algorithms showed that the logistics regression model will provide the highest accuracy of predicting the associated price to tasks. Also, applying PCA wouldn't lead to any better prediction accuracy as data components are not correlated.
The rest of the paper is organized as follows: Section 2 introduces the research design; Section 3 presents the result and discussion of the research conducted; Section 4 gives a conclusion and outlook to future work.

## 2    Research Design

### 2.1    DataSet and Metrics

The dataset used contains 514 component development tasks from Sep 2003 to Sep 2012, extracted from the TopCoder website. All tasks are completed, meaning receiving acceptable submissions with a score higher than 75. The total monetary prize is divided into the top-2 winners with a 2:1 ratio.

The initial analysis on the dataset implies that a typical task on TopCoder is priced as $750 [6] (i.e. $500 and $250 for the top-2 winners respectively), the average size of 2290 lines of code, and the median numbers of registrations and submissions are 16 and 4 respectively. And the median score of the winning submission is 94.16 out of 100. A current common impression is that crowdsourcing is more feasible for easy and simple tasks, the data shows that it is also feasible for complex component development at the scale of 21925 lines of code. The maximum number of registrants of 72 is surprising since the nature of the task is competitive considering that only top-2 winner gets paid.

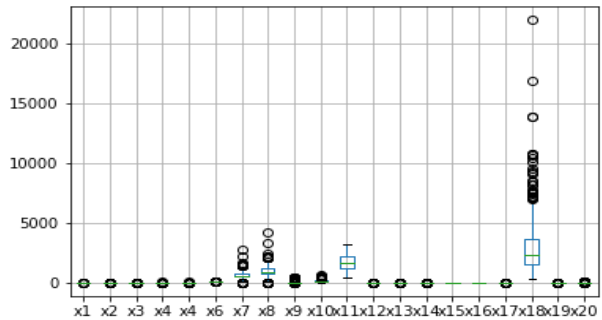

**Fig. 1.** Overall Distribiution of the Dataset

### 2.2    Dataset Preparation

The outlier for each variable was identified and removed. To do so we applied 1.5 of interquartile to add and subtract to the third and first quartiles respectively. Figure 1 illustrates the distribution of data.

Also deeper analysis in the database, it became clear that 8 variables have the least relation with the monetary prize, therefor we ignored them in our analysis. Table 1 summarized the remaining data statistics in the clean database.

Table 1. Summary of metric data in the whole dataset

| Metric | Min | Max | Median | Average | STDEV |
|---|---|---|---|---|---|
| Monetary Prize | 112.5 | 3000 | 750 | 754 | 372.4 |
| Size | 310 | 21925 | 2290 | 2978 | 2267.9 |
| #Reg | 1 | 72 | 16 | 18 | 11 |
| #Sub | 1 | 44 | 4 | 5 | 4 |
| Score | 75 | 100 | 94.1 | 92.5 | 6.2 |

### 2.3 Empirical Studies and Design

In order to predict the monetary prize associated with each task this research studied three predictive modeling methods: linear regression, logistic regression, and KNN.

**Multilinear Regression**
The analysis on the training data suggested that monetary prize ($MP_i$) follows a multiple linear regression, Eq-1:

$$MP_i = \alpha + \sum_i^{20} \beta_i . X_i \qquad (1)$$

In which α is the constant term and $\beta_i$ is the coefficient for the variable ($X_i$).

**Logistic Regression**
As it is reported the average monetary prize for a task in TopCoder is 750$. Therefor to apply logistics regression, we assigned 0 to tasks with a monetary prize of less than 750$ and 1 to tasks with a monetary prize of more than 750$.

**KNN**
To apply KNN, the first 20 different numbers for K were used to find the best neighbor numbers on the clean dataset. Then the PCA method applied to the dataset and the KNN was rerun to study the effect of PCA on the accuracy of KNN results.

## 3    Result and Discussion

### 3.1    Multilinear Regression

Monetary prize is a continuous number, ranging from 75 to 2000. This suggests that the multi linear regression model would be a good match for predicting the monetary prize. Because of this, we first used a multiple regression model to analyze the data. The initial result provides the R2 value of 0.5448 and degree of freedom on 376, which indicates a decent fitting line passing through the dataset. However, a significant number of task variables provide no significant impact on the prediction. Therefore, the insignificant variables were removed and the model re-run. The second model provides the R2 value of 0.5108 and the degree of freedom on 389, which is a worse fit to use for Monetary prize prediction.

The multilinear regression model is created based on 70% training and 30% testing data. The accuracy of the model is shown in figure 2. As it is clear, the prediction model tended to underestimate the monetary prize, however, the median of both actual and predicted monetary prize remained similar.

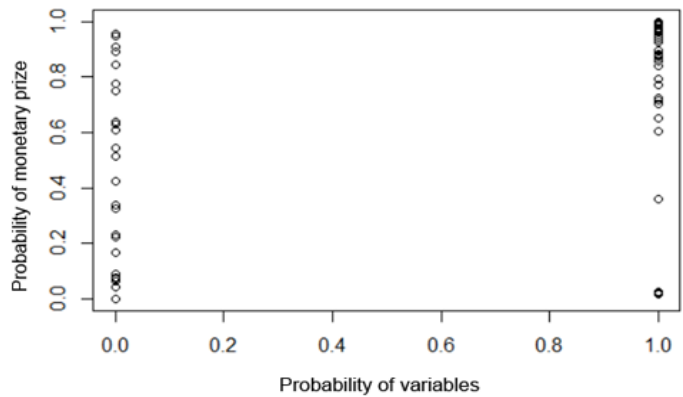

**Fig. 2.** Predicted Monetary Prize by Multi Linear Regression

### 3.2    Logistic Regression

To apply logistic regression to our dataset, the first step is to convert monetary prize to discrete data. Therefore we assigned 0 to price less than 750$ and 1 to price more than 750$. This allowed applying logistics regression to the dataset. Similar to multiple linear regression, the initial model showed that a significant number of task variables were not influential in the equation. Therefore, to make a more efficient model, we only used the significant variables to create the prediction model. Figure 3 presents the classification of Task variables v.s probabilistic prediction of the monetary prize by the logistics regression model. A natural S-Curve shape that the model tends to take is clear.

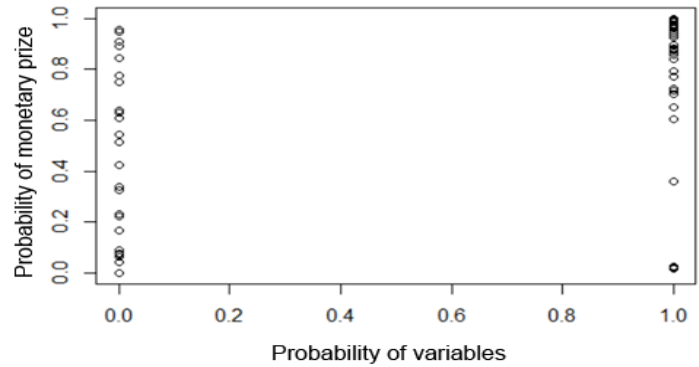

**Fig. 3.** Classification of Variables v.s. Probebilistics on Monetary Prize

Moreover, the confusion matrix of the model with the threshold for prediction at 50% provides a promising result, table 2. The presented model successfully predicts the assigned monetary prize with an accuracy of 90%.

**Table 2.** Confusion Metrix for Logistics Regression Model

| Metric | False | True |
|---|---|---|
| 0 | 16 | 4 |
| 1 | 5 | 68 |

### 3.3  KNN

To find the best K nearest neighbor, the algorithm was run for K from K = 1, to K = 40. Interestingly K = 7 and K = 8 provide the maximum accuracy of 64%. figure 4 represents the error trend for K between 1 to 40. And table 3 reports the classification report for K = 7 as an example.

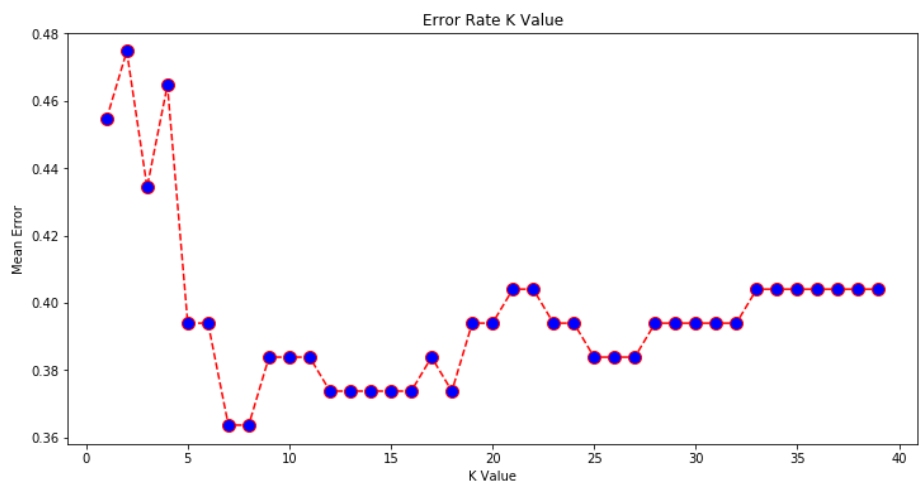

**Fig. 4.** Error Trend for K between 1 to 40

**Table 3.** Classification Report of K = 7, KNN Model

|      | precision | recall | f1-score | support |
|------|-----------|--------|----------|---------|
| 75   | 0         | 0      | 0        | 0       |
| 96   | 0.67      | 1      | 0.8      | 2       |
| 100  | 0.67      | 0.67   | 0.67     | 3       |
| 144  | 0         | 0      | 0        | 1       |
| 150  | 0         | 0      | 0        | 3       |
| 250  | 0         | 0      | 0        | 3       |
| 300  | 0.33      | 0.5    | 0.4      | 2       |
| 400  | 0         | 0      | 0        | 7       |
| 500  | 0         | 0      | 0        | 1       |
| 600  | 0         | 0      | 0        | 4       |
| 700  | 0         | 0      | 0        | 4       |
| 750  | 0.67      | 0.98   | 0.8      | 59      |
| 800  | 0         | 0      | 0        | 2       |
| 900  | 0         | 0      | 0        | 1       |
| 1000 | 0         | 0      | 0        | 4       |
| 1250 | 0         | 0      | 0        | 1       |
| 1500 | 0         | 0      | 0        | 1       |
| 1800 | 0         | 0      | 0        | 1       |
|      |           |        |          |         |
| micro avg    | 0.64 | 0.64 | 0.64 | 99 |
| macro avg    | 0.13 | 0.17 | 0.15 | 99 |
| weighted avg | 0.44 | 0.64 | 0.52 | 99 |

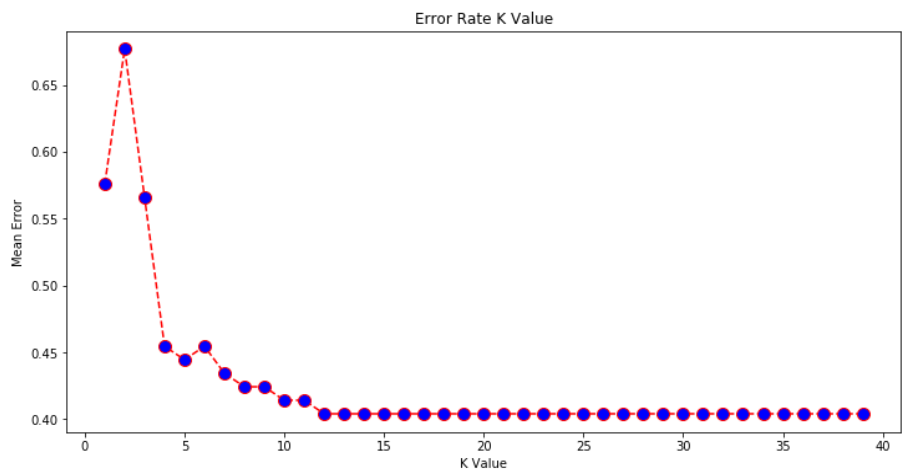

**Fig. 5.** Error Trend for K between 1 to 40 after Applying PCA

In the next step, the PCA was applied and the accuracy of KNN under PCA was analyzed. The result showed that PCA not only could not improve accuracy but also it decreased by 4%. Figure 5 shows the error rate for the KNN model after applying PCA for K between 1 to 40.

## 4       Conclusions

The monetary prize usually represents the degree of task complexity as well as required competition levels. In practice, task requesters frequently employ expert-based methods to price tasks, which may involve a high degree of subjectivity, while the challenge of validating the associated prices across different tasks is a constant issue.

To address this issue, three different classifications of multiple linear regression, logistic regression, and K-nearest neighbor were compared to find the most accurate predicted price, using a dataset from the TopCoder website. The result of comparing chosen algorithms showed that the logistics regression model will provide the highest accuracy of 90% to predict the associated price to tasks and KNN ranked the second with an accuracy of 64% for K = 7. Also, applying PCA wouldn't lead to any better prediction accuracy as data components are not correlated.

In the future, we would like to focus on the similar crowd worker behavior and performance based on task similarity level and try to analyze a task- worker performance to report more decision elements according to the monetary prize, task size task utilization, and crowd workers' performance.